\documentclass[final,3p,times]{elsarticle}

\usepackage{CJK}

\usepackage{graphicx} 
\usepackage{amssymb}
\usepackage{amsmath}
\usepackage{subfigure}
\usepackage{graphicx}
\usepackage{epstopdf}
\usepackage{bibentry}
\usepackage{color}
\usepackage{breqn}
\usepackage{lineno,hyperref}

\allowdisplaybreaks[4]

\journal{}

\begin{document}

\begin{frontmatter}



\title{{Two $(2+1)$-dimensional integrable nonlocal nonlinear Schr\"{o}dinger equations: Breather, rational and semi-rational solutions}}



\author[1]{Yulei Cao}
\author[2,3]{Boris A. Malomed}
\author[1]{\corref{cor1} Jingsong He}
\address[1]{Department of Mathematics, Ningbo University,
Ningbo, Zhejiang, 315211, P.\ R.\ China}
\address[2]{Department of Physical Electronics, Faculty of Engineering,
Tel Aviv University, Tel Aviv 69978, Israel }
\address[3]{Laboratory of Nonlinear-Optical Informatics, ITMO University,
St. Petersburg 197101, Russia}
\cortext[cor1]{e-mail: hejingsong@nbu.edu.cn;  jshe@ustc.edu.cn}

\begin{abstract}
Recently, an integrable system of coupled $(2+1)$-dimensional nonlinear Schr%
\"{o}dinger (NLS) equations was introduced by Fokas (eq. (18) in
Nonlinearity \textbf{29}, 319324 (2016)). Following this pattern, two
integrable equations [eqs.\eqref{N1} and \eqref{N2}] with specific
parity-time symmetry are introduced here, under different reduction
conditions. For eq. \eqref{N1}, two kinds of periodic solutions are obtained
analytically by means of the Hirota's bilinear method. In the long-wave
limit, the two periodic solutions go over into rogue waves (RWs) and
semi-rational solutions, respectively. The RWs have a line shape, while the
semi-rational states represent RWs built on top of the background of
periodic line waves. Similarly, semi-rational solutions consisting of a line
RW and line breather are derived. For eq. \eqref{N2}, three kinds of
analytical solutions,\textit{viz}., breathers, lumps and semi-rational
solutions, representing lumps, periodic line waves and breathers are
obtained, using the Hirota method. Their dynamics are analyzed and
demonstrated by means of three-dimensional plots. It is also worthy to note
that eq. \eqref{N1} can reduce to a $(1+1)$-dimensional \textquotedblleft
reverse-space" nonlocal NLS equation by means of a certain transformation,
Lastly, main differences between solutions of eqs.\eqref{N1} and \eqref{N2}
are summarized.
\end{abstract}


\end{frontmatter}


\section{Introduction}
Rogue waves (RWs) have become famous due to their realization as
spontaneously emerging freak surface waves in the ocean (where they are held
responsible for some maritime disasters). In this context, they are often
characterized as localized patterns which \textquotedblleft appear from
nowhere and then disappear without a trace" \cite{AAM-1}. RWs have been
identified in a number of other contexts, including Bose-Einstein
condensates \cite{BKA1,BKA2}, optical systems \cite{MBR1,MBR2}, oceanography
\cite{KPS1}, plasmas \cite{MMW1,MMW2}, etc. The Peregrine soliton is the
fundamental (first-order) RW solution produced by the one-dimensional ($1$D)
nonlinear Schr\"{o}dinger (NLS) equation \cite{PHD1}. Lately, higher-order
RW solutions of the NLS equation have been produced by different methods
\cite{OY1,OY2,OY3,OY4,OY5,OY6,OY7}. Moreover, a series of other soliton equations have
also been shown to possess RW solutions \cite{PD1,PD2,PD3,PD4,PD5,PD6,PD7,PD8}. Detailed
reviews of theoretical and experimental aspects of RWs are available in
Refs. \cite{onorato, NatureReview, Onorato2016}. Due to the fact that the
ocean surface is two-dimensional ($2$D), studies of RWs have been naturally
extended from $1$D to 2D models. In particular, RW solutions of several
important integrable $2$D equations have been derived explicitly, such as
the Davey-Stewartson (DS) \cite{OYY1,OYY2} and Kadomtsev-Petviashvili-I (KP
I) \cite{NatureReview,DDE2} equations, and others \cite{MQ1,MQ2,CJ1,CJ2}.
Similar solutions were found even in the $3$D KP equation \cite{3DKP}.

It is relevant to mention that RWs are essentially unstable solutions, as
they are supported by the modulationally unstable background. An interesting
possibility is to effectively stabilize RW-like modes by means of the
nonlinearity- and dispersion- \textquotedblleft management" techniques \cite%
{Cuevas}, even if the integrability of the underlying NLS equation is lost
in such a setting.

Recently, Fokas \cite{FAS1} has found an integrable system of coupled $2$D NLS
equations,
\begin{equation}
\begin{aligned}
iU_{t}+U_{xy}-\frac{1}{2}(\partial_{z}^{-1}+\partial_{%
\bar{z}}^{-1})U[UV]_{y}=0,\\
iV_{t}-V_{xy}+\frac{1}{2}(\partial_{z}^{-1}+\partial_{%
\bar{z}}^{-1})V[UV]_{y}=0,\\ \end{aligned}  \label{eq1}
\end{equation}%
obtained as a special reduction to a more general $(4+2)$-dimensional system
\cite{FAS2}; here, $z\equiv x+iy$, and $\partial _{z,\bar{z}}^{-1}$ are
operators inverse to $\partial _{z}\equiv (1/2)\left( \partial
_{x}-i\partial _{y}\right) $ and $\partial _{\bar{z}}\equiv (1/2)\left(
\partial _{x}+i\partial _{y}\right) $; in particular, $\partial _{\bar{z}%
}^{-1}=\int \int \left( z-z^{\prime }\right) ^{-1}f\left( x^{\prime
},y^{\prime }\right) dx^{\prime }dy^{\prime }$. Here, $U$ and $V$ are
complex functions of $x,y$ and $t$.

If $y=x$, then $\partial _{z}^{-1}+\partial _{\bar{z}}^{-1}=\partial
_{x}^{-1}$ and eq. \eqref{eq1} reduces to the classical system of AKNS
equations \cite{AKNS}. Eq. \eqref{eq1} becomes the following integrable
\textquotedblleft reverse-space" $2$D nonlocal NLS equation \cite{FAS1} when
$V(x,y,t)=\lambda U^{\ast }(-x,-y,t)$, and $\lambda =\pm 1$:
\begin{equation}
(i\partial _{t}+\partial _{xy}^{2})U(x,y,t)-\frac{\lambda }{2}%
U(x,y,t)(\partial _{z}^{-1}+\partial _{\bar{z}}^{-1})\partial
_{y}[U(x,y,t)U^{\ast }(-x,-y,t)]=0,\lambda =\pm 1,  \label{N1}
\end{equation}%
If $V(x,y,t)=\lambda U^{\ast }(-x,y,-t),\lambda =\pm 1$, a reverse
space-time nonlocal NLS equation is introduced:
\begin{equation}
(i\partial _{t}+\partial _{xy}^{2})U(x,y,t)-\frac{\lambda }{2}%
U(x,y,t)(\partial _{z}^{-1}+\partial _{\bar{z}}^{-1})\partial
_{y}[U(x,y,t)U^{\ast }(-x,y,-t)]=0,\lambda =\pm 1,  \label{N2}
\end{equation}

Eqs. \eqref{N1} and \eqref{N2} satisfy the condition of the $2$D parity-time
($\mathcal{PT}$) symmetry. The concept of the $\mathcal{PT}$ symmetry in the
quantum theory has been introduced in pioneering works which aimed to
construct non-Hermitian Hamiltonians with real spectra \cite{BBS1,BBS2}.
Various states generated by $\mathcal{PT}$-symmetric systems are supported
by the balance of loss and gain inherent to this symmetry \cite{MGC1,MGC2},
see comprehensive review \cite{KYZ1}. Many results for nonlinear
$\mathcal{PT}$-symmetric models, where the gain-loss balance is
combined with the usual equilibrium between diffraction and nonlinear
self-interaction, were summarized in reviews \cite{KYZ1} and \cite{par5}.\
The $\mathcal{PT}$-symmetry has also
been studied extensively in multi-dimensional systems \cite{par5, yang,
par1,par2,par3,par4}. An issue of obvious interest is to produce
multi-dimensional integrable models featuring the $\mathcal{PT}$ symmetry,
which is a motivation for the consideration of eqs. \eqref{N1} and \eqref{N2}%
, and to explore dynamics of RWs in $\mathcal{PT}$-symmetry.

The rest of the paper is organized as follows. In sec. \ref{2}, periodic
line-wave, RW, and semi-rational solutions, composed of first-order RWs,
line breathers and periodic line waves of eq. \eqref{N1} are produced by
means of the Hirota method. In the same section, typical dynamics of these
solutions are analyzed and demonstrated. In sec. \ref{3}, three kinds of
analytical solutions of eq. \eqref{N2}, namely, breathers, lumps and
semi-rational solutions, consisting of lump modes, periodic line waves, and
breathers are derived and illustrated. The main results of the paper are
summarized in sec. \ref{5}.


\section{Solutions of the reverse-space nonlocal $2$D NLS equation(\protect
\ref{N1})}

\label{2}

In this section, we focus on eq. (\ref{N1}). Setting
\begin{equation}
\begin{aligned} V_{x}=[U(x,y,t)U^{*}(-x,-y,t)]_{y},\\ \end{aligned}
\label{T1}
\end{equation}%
Eq. \eqref{N1} becomes the following system of coupled partial differential
equations:
\begin{equation}
\begin{aligned} &iU_{t}+U_{xy}+UV=0,\\
&V_{x}=[U(x,y,t)U^{*}(-x,-y,t)]_{y},\\ \end{aligned}  \label{D1}
\end{equation}%
Here $U$ and $V$ are two complex functions of $x,y$ and $t$, and $V$
satisfies the two-dimensional PT symmetry condition $V(x,y,t)=V^{\ast
}(-x,-y,t)$. This observation inspires us to study \eqref{N1} or equivalent %
\eqref{D1}. employing the bilinear method. To this end, the equation can be
translated into the following bilinear form:
\begin{gather}
(D_{x}D_{y}+iD_{t})g\cdot f=0,  \notag \\
(D_{x}^{2}+1)f\cdot f=gg^{\ast }(-x,-y,t),  \label{N4}
\end{gather}%
through the transformation of the dependent variables:
\begin{equation}
U=g/f,\qquad V=2(\ln f)_{xy},  \label{N5}
\end{equation}%
where $f$ and $g$ are functions of $x$, $y$ and $t$, subject to the
condition:
\begin{equation}
f(x,y,t)=f^{\ast }(-x,-y,t).  \label{N6}
\end{equation}%
$D$ is the Hirota's bilinear differential operator \cite{hirota} defined by
\begin{gather}
P(D_{x},D_{y},D_{t},)F(x,y,t\cdot \cdot \cdot )\cdot G(x,y,t,\cdot \cdot
\cdot ) \\
=P(\partial _{x}-\partial _{x^{^{\prime }}},\partial _{y}-\partial
_{y^{^{\prime }}},\partial _{t}-\partial _{t^{^{\prime }}},\cdot \cdot \cdot
)F(x,y,t,\cdot \cdot \cdot )G(x^{^{\prime }},y^{^{\prime }},t^{^{\prime
}},\cdot \cdot \cdot )|_{x^{^{\prime }}=x,y^{^{\prime }}=y,t^{^{\prime }}=t},
\end{gather}%
where $P$ is a polynomial of $D_{x}$,$D_{y}$,$D_{t},\cdot \cdot \cdot $.

The N-soliton solutions $U$ and $V$ given in \eqref{N5} of the eq. \eqref{N1}
can be obtained by the bilinear transform method \cite{hirota}, in which $f$
and $g$ are written in the following forms:
\begin{equation}  \label{N7}
\begin{aligned}
f=\sum_{\mu=0,1}\exp(\sum_{j<k}^{(N)}\mu_{j}\mu_{k}A_{jk}+\sum_{j=1}^{N}%
\mu_{j}\eta_{j}),\; \;
g=\sum_{\mu=0,1}\exp(\sum_{j<k}^{(N)}\mu_{j}\mu_{k}A_{jk}+\sum_{j=1}^{N}%
\mu_{j}(\eta_{j}+i\Phi_{j})). \end{aligned}
\end{equation}
Here
\begin{equation}  \label{N8}
\begin{aligned} \Omega _{j}=-Q_{j}\sqrt{-P_{j}^{2}+2}, \eta
_{j}=iP_{j}x+iQ_{j}y+\Omega _{j}t+\eta _{j}^{0}, \cos \Phi
_{j}=-P_{j}^{2}+1, \\ \sin \Phi _{j}=\sqrt{(-P_{j}^{2}+2)}P_{j}, \exp
(A_{jk})=-\frac{\cos (\Phi _{j}-\Phi _{k})+ (P_{j}-P_{k})^{2}-1}{\cos (\Phi
_{j}+\Phi _{k})+(P_{j}+P_{k})^{2}-1}, \end{aligned}
\end{equation}
where $P_{j},Q_{j},$ are arbitrary real parameters, and $\eta _{j}^{0}$ are
complex constants. The notation $\sum_{\mu =0,1}$ implies summation over all
possible combinations of $\mu _{1}=0,1,\mu _{2}=0,1,\cdots ,\mu _{n}=0,1$.
The $\sum\limits_{j<k}^{N}$ summation is performed over all possible
combinations of the $N$ elements, subject to condition $j<k$.
\subsection{periodic line wave solutions}

\begin{figure}[tbh]
\centering
\subfigure[t=-8]{\includegraphics[height=3cm,width=5.0cm]{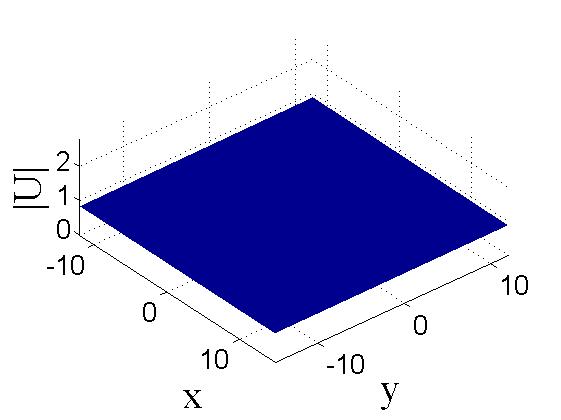}} %
\subfigure[t=-0.5]{\includegraphics[height=3cm,width=5.0cm]{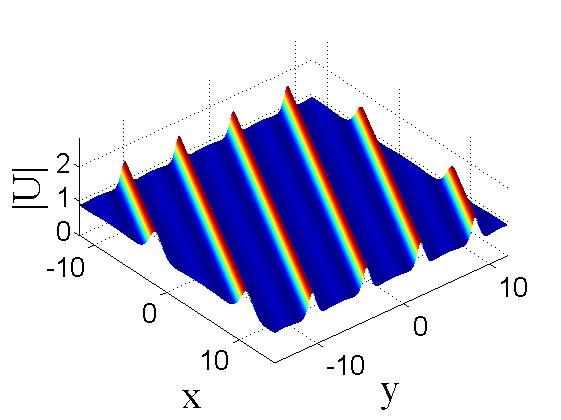}} %
\subfigure[t=0]{\includegraphics[height=3cm,width=5.0cm]{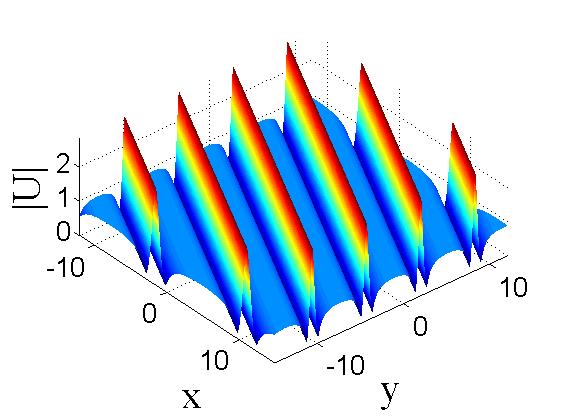}} %
\subfigure[t=2]{\includegraphics[height=3cm,width=5.0cm]{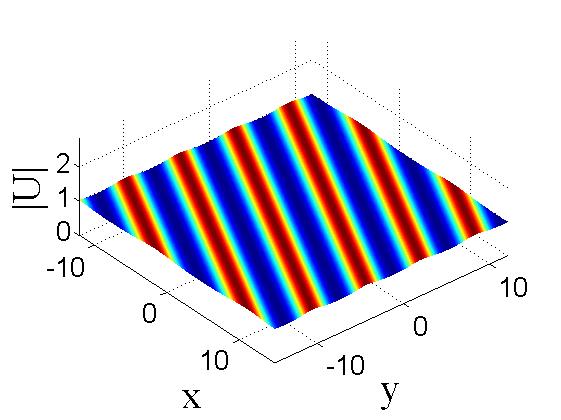}} %
\subfigure[t=8]{\includegraphics[height=3cm,width=5.0cm]{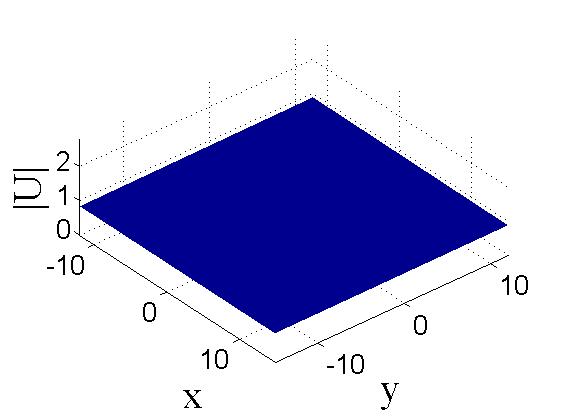}}
\caption{The evolution of the first-order line breather of eq. \eqref{N1} in
the $(x,y)$-plane, corresponding to parameters $P_{1}=\frac{1}{2}$ and $%
Q_{1}=1$ in eq. (\protect\ref{7Nc}).}
\label{fig1}
\end{figure}

Following previous works \cite{rao1,rao2,rao3,rao4}, $n$-th-order line breather
solutions of the eq. \eqref{N1} can be generated by taking parameters in eq.%
\eqref{N7} as
\begin{equation}
N=2n,P_{j}=-P_{n+j},Q_{j}=-Q_{n+j},\eta _{j}^{0}=\eta _{n+j}^{0}.
\label{N7b}
\end{equation}%
For instance, a particular case of eq. (\ref{N7b}), with
\begin{equation}
N=2,P_{1}=-P_{2},Q_{1}=-Q_{2},\eta _{1}^{0}=\eta _{2}^{0}=0,  \label{7Nc}
\end{equation}%
produces the first-order breather solution which is shown in Fig. \ref{fig1}%
. It is seen that the corresponding solution describes a periodic array of
line waves in the $(x,y)$ plane, first growing and then decaying in time.
Five panels in Fig. \ref{fig1} show that the first-order line breather
appears at $t<0$ from a flat background, gradually attains a maximum
amplitude at $t=0$, and finally returns back to the asymptotic background
without a trace, which is typical to RWs.

\begin{figure}[tbh]
\centering
\subfigure[t=-20]{\includegraphics[height=3cm,width=3.7cm]{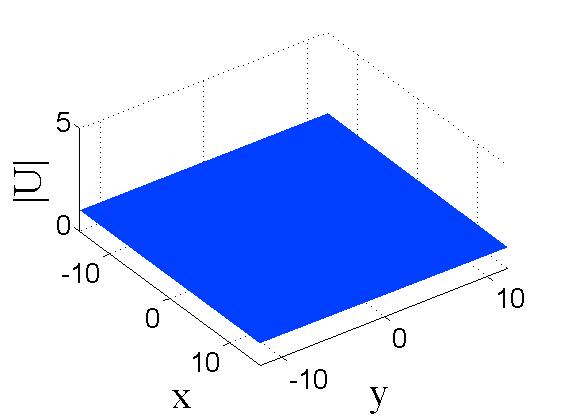}} %
\subfigure[t=-16]{\includegraphics[height=3cm,width=3.7cm]{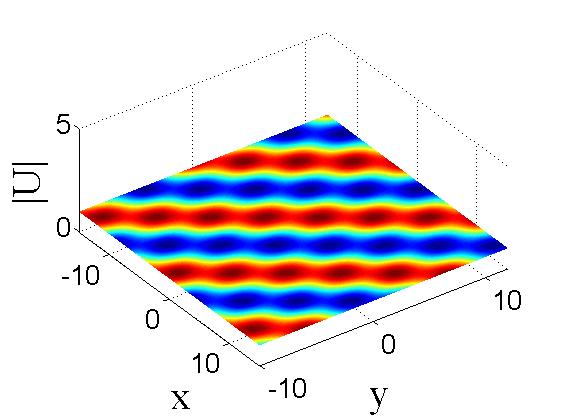}} %
\subfigure[t=-1]{\includegraphics[height=3cm,width=3.7cm]{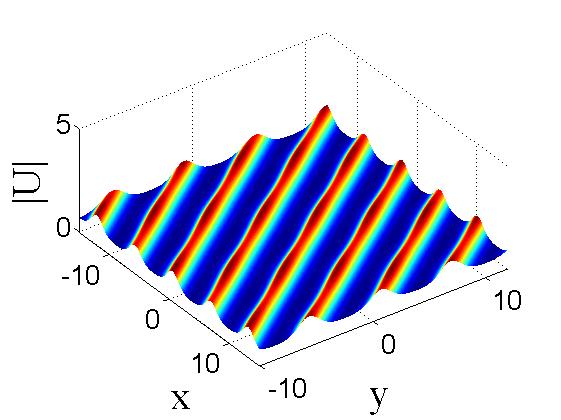}} %
\subfigure[t=0]{\includegraphics[height=3cm,width=3.7cm]{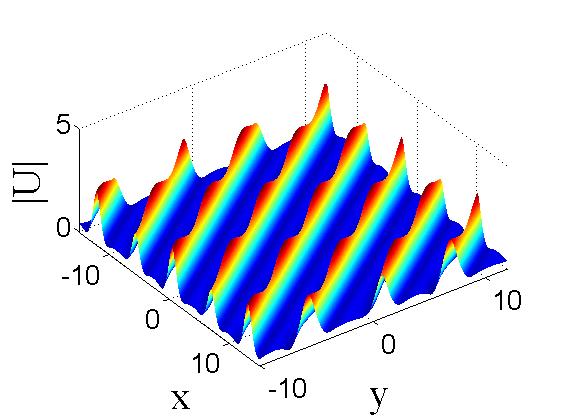}} %
\subfigure[t=2]{\includegraphics[height=3cm,width=3.8cm]{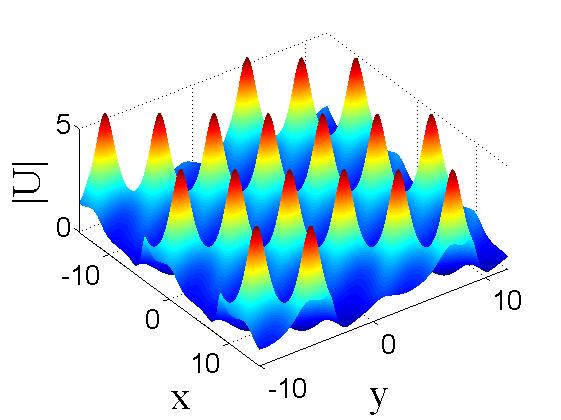}} %
\subfigure[t=3]{\includegraphics[height=3cm,width=3.7cm]{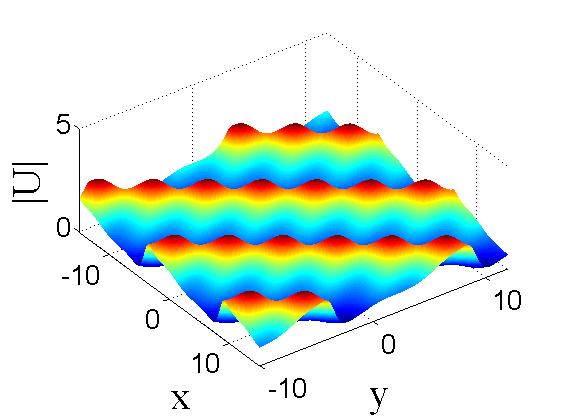}} %
\subfigure[t=4]{\includegraphics[height=3cm,width=3.7cm]{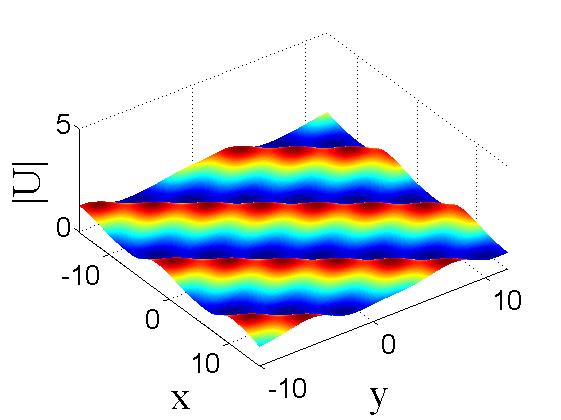}} %
\subfigure[t=8]{\includegraphics[height=3cm,width=3.7cm]{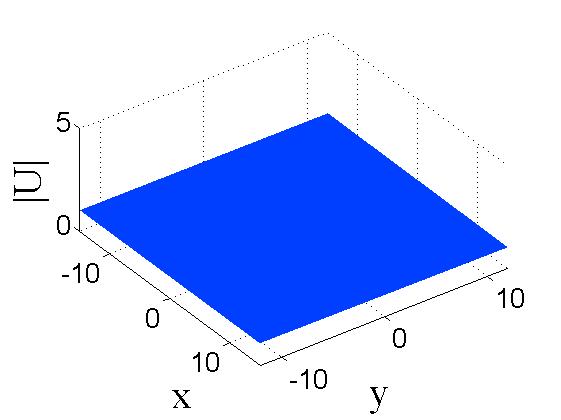}}
\caption{The evolution of the third-order periodic solution $|U|$ of eq. (%
\protect\ref{N1}) in the $(x,y)$-plane, with parameters $%
P_{1}=1,Q_{1}=1,P_{3}=\frac{1}{2}$, and $\protect\eta _{1}^{0}=0,\protect%
\eta _{3}^{0}=-\frac{\protect\pi }{2}$ in eq. (\protect\ref{7Ne}).}
\label{fig2}
\end{figure}

Additionally, for the case of $N=2n+1$, another type of periodic solutions
composed of line breathers and periodic line waves can be generated via eq. %
\eqref{N5}, taking parameters
\begin{equation}
N=2n+1,P_{j}=-P_{n+j},Q_{j}=-Q_{n+j},\eta _{j}^{0}=\eta
_{n+j}^{0},P_{2n+1}Q_{2n+1}\neq0  \label{7Nd}
\end{equation}%
in eq. \eqref{N7}. For example, setting
\begin{equation}
N=3,P_{1}=-P_{2},Q_{1}=-Q_{2},P_{3}=-Q_{3},\eta _{1}^{0}=\eta
_{2}^{0},P_{3}Q_{3}\neq 0,  \label{7Ne}
\end{equation}%
in eq. (\ref{7Nd}), we obtain a mixed solution composed of the first-order
line breather and first-order periodic line wave. The dynamics of the
emergence and disappearance of this third-order solution is shown in Fig. %
\ref{fig2}. It is seen that the first-order periodic line wave arises from
the constant background at $t<-16$, which possesses one maximum and one
minimum amplitude, featuring a clear parallel-line profile in the horizontal
direction(see Fig.\ref{fig2}b). Subsequently, the first-order line breather
appears in the vertical direction, overlaid on the periodic line wave around
$t=-1$. Due to the amplitude of the periodic line waves are too small, so
the interaction phenomenon of periodic line waves and line breather are
covered up. Along the evolution in time the characteristic of the line
breather becomes clearer, which has one maximum and two minimum amplitudes
(See Fig.\ref{fig2}d, it is similar to Fig.\ref{fig1}c), and the phenomenon
of interaction is also gradually obvious at $t=0$, then an array of sharp
peaks is observed at $t=2$, produced by the interplay of the two wave
patterns. Next, the first-order line breather quickly merges back into
background, but the first-order periodic line wave is observed around $t=4$.
Finally, this periodic line-wave pattern disappears at $t>8$. Note that this
2D dynamical solution is different from the one obtained by means of the
same method in the nonlocal DS I and II equations \cite{rao1,rao2}. Further,
it is relevant to mention that Fig. \ref{fig2}(d) is similar to the doubly
periodic line wave in the $(x,t)$ (rather than $\left( x,y\right) $) plane,
obtained in \cite{fan} (see Fig.1(c) in that work).

\subsection{Rational and semi-rational solutions.}

To generate RW solutions of eq. (\ref{N1}), the long-wave limit of
thebreather solutions given by eq. \eqref{N7} may be used. Indeed, taking
parameters in eq. \eqref{N7} as
\begin{equation}
N=2,Q_{1}=\lambda _{1}P_{1},Q_{2}=\lambda _{2}P_{2},\lambda _{1}=\lambda
_{2}=\lambda\neq0,  \label{N9}
\end{equation}%
and taking the limit of $P_{j}\rightarrow 0$ $(j=1,2)$, the first-order RW
solution is obtained as
\begin{eqnarray}
U &=&\frac{(\theta _{1}+b_{1})(\theta _{2}+b_{2})+\alpha _{12}}{\theta
_{1}\theta _{2}+\alpha _{12}}=\frac{2(\lambda y+x)^{2}+4(\lambda t-i)^{2}+1}{%
2(\lambda y+x)^{2}+4(\lambda t)^{2}+1},  \notag \\
V &=&2(\ln (\theta _{1}\theta _{2}+\alpha _{12})_{xy}=\frac{2\lambda
_{1}[-2(\lambda y+x)^{2}+4(\lambda t)^{2}+1]}{[(\lambda y+x)^{2})+2(\lambda
t)^{2}+\frac{1}{2}]^{2}},  \label{N10}
\end{eqnarray}%
where
\begin{eqnarray}
&&\theta _{j}=i\lambda _{j}y-\sqrt{2}\lambda _{j}t+ix, b_{1}=-b_{2}=i\sqrt{2}%
, \alpha _{12}=-\frac{1}{2},  \label{N11}
\end{eqnarray}%
with the corresponding profile of $|U\left( x,y\right) |$ shown in Fig.\ref%
{fig3}. It is seen that this W-shaped solution describes an emerging and
decaying line wave oriented in the $(\lambda ,-1)$ direction of the $(x,y)$%
-plane. At any given time, this solution keeps a constant value along the
line direction defined by $\lambda y+x=0$, and $|U\left( x,y\right) | $
uniformly approaches the flat background at $t\rightarrow \pm \infty $. At
finite times, $|U\left( x,y\right) |$ attains a maximum at the center ($%
\lambda y+x=0$) of the line wave. In particular, the maximum value is $3$
(i.e., three times the background amplitude) at $t=0$. We stress that the
rational solution given by Eq. (\ref{N10}) can not be directly reduced to
rational solutions of the $\mathcal{PT}$ symmetric DS I equation found in
Ref. \cite{rao1,rao2}.
\begin{figure}[tbh]
\centering
\subfigure[t=-3]{\includegraphics[height=3cm,width=5cm]{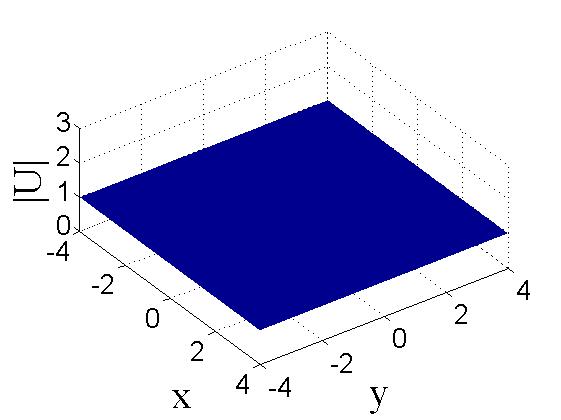}} %
\subfigure[t=-0.25]{\includegraphics[height=3cm,width=5cm]{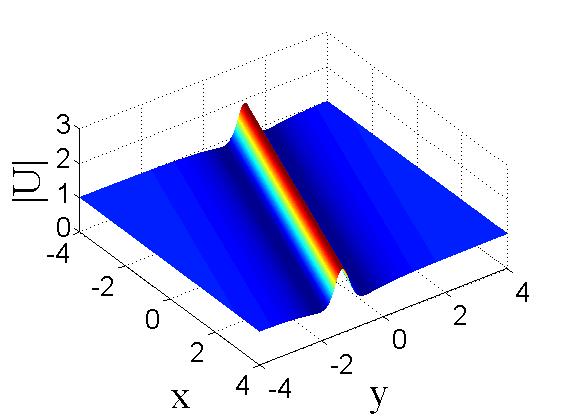}} %
\subfigure[t=0]{\includegraphics[height=3cm,width=5cm]{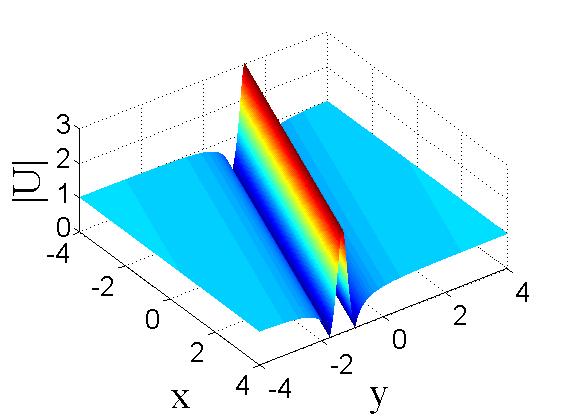}} %
\subfigure[t=0.5]{\includegraphics[height=3cm,width=5cm]{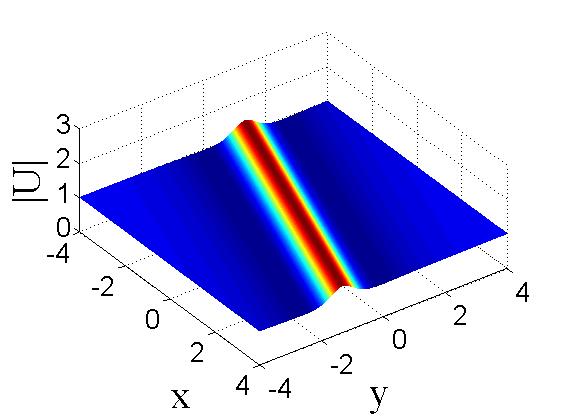}} %
\subfigure[t=3]{\includegraphics[height=3cm,width=5cm]{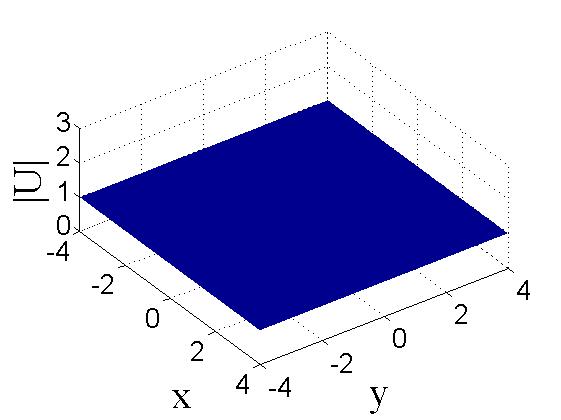}}
\caption{The time evolution of the first-order (W-shaped) line rogue wave in
the $(x,y)$ plane, given by solution (\protect\ref{N10}), (\protect\ref{N11}%
) of eq. (\protect\ref{N1}), with parameter $\protect\lambda =3$ in eq. (%
\protect\ref{N9}). }
\label{fig3}
\end{figure}

Furthermore, semi-rational solutions of eq. (\ref{N1}) can be obtained by
taking the long-wave limit of periodic solutions, which describe the
interplay of RWs with periodic waves. Indeed, setting parameters in eq. %
\eqref{N7} to be
\begin{equation}
0<2j<N,1\leq k\leq 2j,Q_{k}=\lambda _{k}P_{k},\eta _{k}^{0}=i\pi ,
\label{N12}
\end{equation}%
and taking the limit of $P_{k}\rightarrow 0$, functions $f$ and $g$ defined
by eq. \eqref{N7} become a combination of polynomial and exponential
functions, which generate a semi-rational solution for $U\left( x,y,t\right)
$ and $V\left( x,y,t\right) $ of eq. (\ref{N1}) via eq. \eqref{N5}.

To clearly illustrate this method for constructing the semi-rational
solutions, we take here $N=3$ and $\Omega _{3}=0$. Setting
\begin{equation}
Q_{k}=\lambda _{k}P_{k}, \quad \eta _{k}^{0}=i\pi (k=1,2),  \label{N13}
\end{equation}%
and taking the limit of $P_{k}\rightarrow 0$ in eq. \eqref{N7}, we obtain
\begin{eqnarray}
f &=&(\theta _{1}\theta _{2}+a_{12})+(\theta _{1}\theta
_{2}+a_{12}+a_{13}\theta _{2}+a_{23}\theta _{1}+a_{12}a_{23})e^{\eta _{3}},
\notag \\
g &=&[(\theta _{1}+b_{1})(\theta _{2}+b_{2})+a_{12}]+[(\theta
_{1}+b_{1})(\theta _{2}+b_{2})+a_{12}+a_{13}(\theta _{2}+b_{2})  \notag \\
&&+a_{23}(\theta _{1}+b_{1})+a_{12}a_{23})]e^{\eta _{3}+i\phi _{3}},
\label{N14}
\end{eqnarray}%
where $a_{j3}=1-2p_{j}P_{3}\left[ \sqrt{2\left( -P_{3}^{2}+2\right) }+2%
\right] ^{-1}\delta _{j}$, $\delta _{j}=\pm 1$ ($j=1,2$), and $%
a_{12}\,,b_{j}\,,\Phi _{j},\eta _{3}$ are given by eqs. \eqref{N8} and %
\eqref{N11}. The corresponding semi-rational solution is shown in Fig. \ref%
{fig4}. It is seen that it describes a line RW on top of the background of
periodic line waves. When $t\rightarrow \pm \infty $, the line RW approaches
the background, see the panels corresponding to $t=\pm 5$. At the
intermediate time moment, $t=-0.35$, the line RW arises from the periodic
background, then the interplay around $t=0$ between the line RW and the
periodic line-waves background creates sharp peaks along the line RW, which
can reach the maximum value of $8$ (i.e., $8$ times the background
amplitude). In the course of the subsequent evolution, all peaks on the line
RW gradually merge into periodic line waves. Finally, the wave field returns
to the background formed by the periodic line waves at $t\rightarrow \infty $%
. This solution is, in particular, notably different from the corresponding
solutions with the flat background of the nonlocal DS I equation presented
in Ref. \cite{rao1}, see Fig. 12 in that work.
\begin{figure}[tbh]
\centering
\subfigure[t=-5]{\includegraphics[height=3cm,width=4cm]{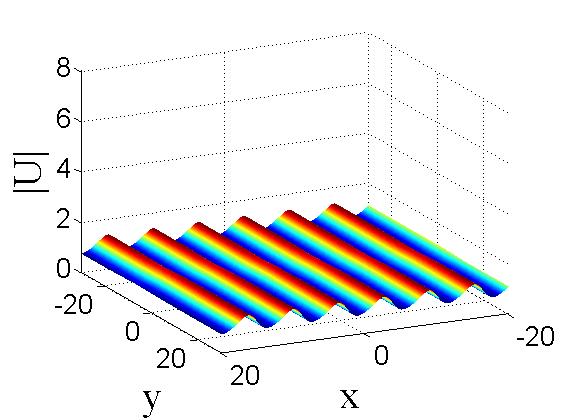}} %
\subfigure[t=-0.35]{\includegraphics[height=3cm,width=4cm]{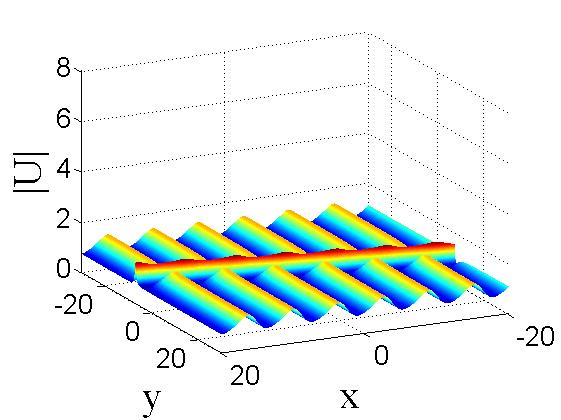}} %
\subfigure[t=-0.15]{\includegraphics[height=3cm,width=4cm]{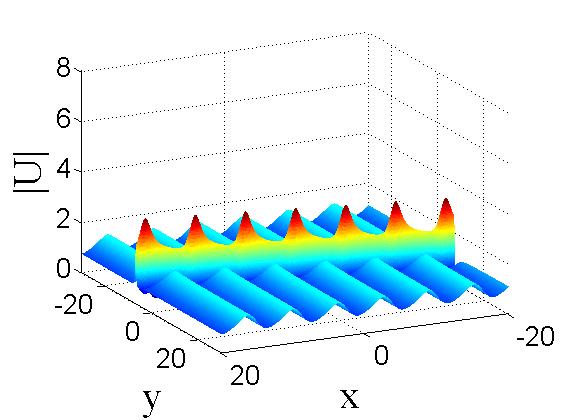}}\newline
\subfigure[t=0]{\includegraphics[height=3cm,width=4cm]{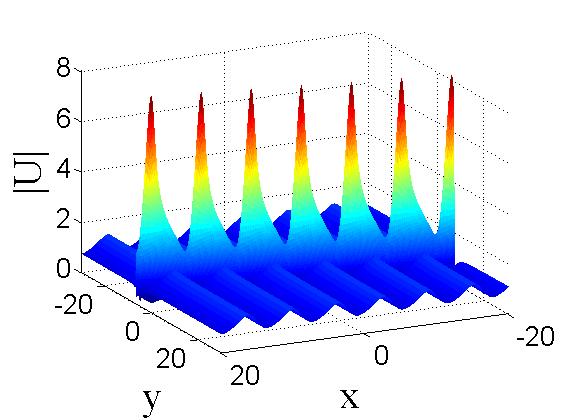}} %
\subfigure[t=0.25]{\includegraphics[height=3cm,width=4cm]{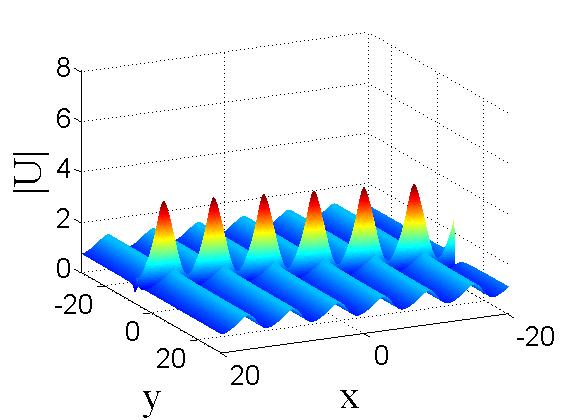}} %
\subfigure[t=5]{\includegraphics[height=3cm,width=4cm]{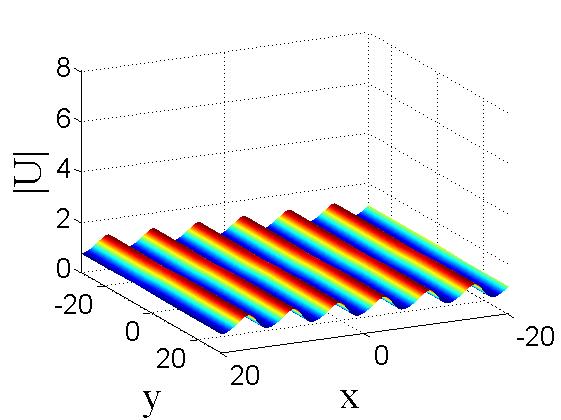}}
\caption{The evolution of the line RW on top of the background formed by
periodic line waves with parameters $\protect\lambda _{1}=\protect\lambda %
_{2}=3,Q_{3}=0,P_{3}=1$, and $\protect\eta _{3}^{0}=-\protect\pi /2$ in eq. (%
\protect\ref{N13}).}
\label{fig4}
\end{figure}

\begin{figure}[!htb]
\centering
\subfigure[t=-5]{\includegraphics[height=3cm,width=5cm]{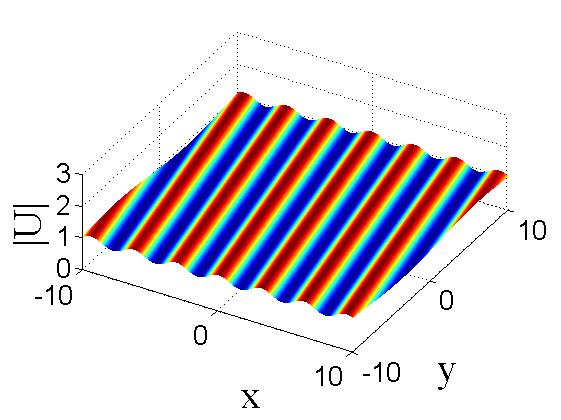}} %
\subfigure[t=-1]{\includegraphics[height=3cm,width=5cm]{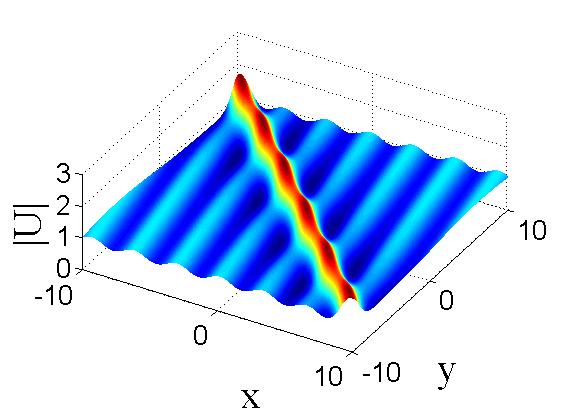}} %
\subfigure[t=0]{\includegraphics[height=3cm,width=5cm]{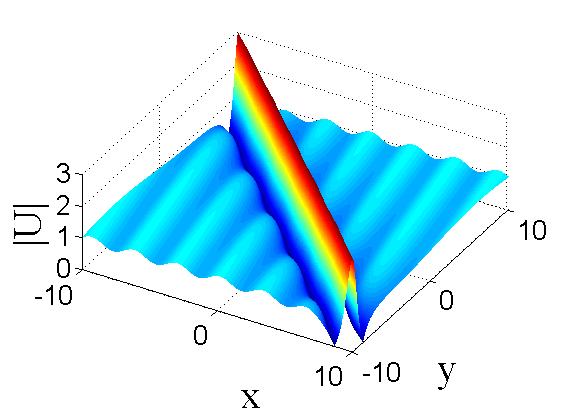}} %
\subfigure[t=0.5]{\includegraphics[height=3cm,width=5cm]{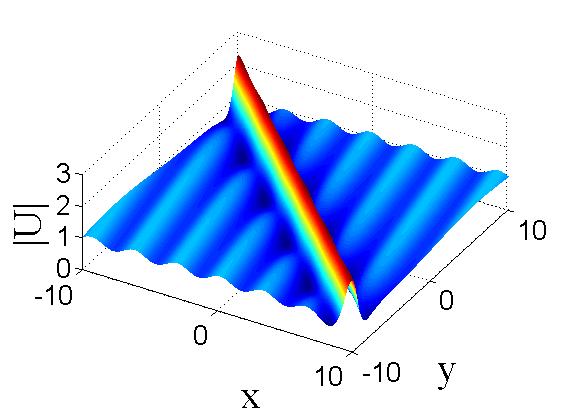}} %
\subfigure[t=5]{\includegraphics[height=3cm,width=5cm]{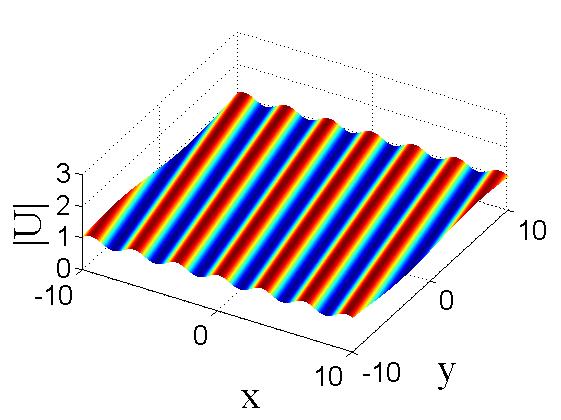}}
\caption{The evolution of the solution to eq. \eqref{N1} in the form of
the line rogue wave, built on top of the line-breather
background, in the $(x,y)$-plane}
\label{rb}
\end{figure}

Furthermore, higher-order semi-rational solutions consisting of RW and line
breather solutions can also be generated in a similar way for larger values
of $N$. We have to take the parameters
\begin{equation}  \label{Nrl1}
\begin{aligned}
N=4\,,Q_{1}=\lambda_{1}P_{1}\,,Q_{2}=\lambda_{2}P_{2}\,,\eta_{1}^{0}=%
\eta_{2}^{0}=i\,\pi\,,\eta_{3}^{0}=\eta_{4}^{0}, \end{aligned}
\end{equation}
and take a limit as $P_{1}\,,P_{2}\rightarrow 0$. Then the functions $f$ and
$g$ can be rewritten as
\begin{equation}  \label{rl2}
\begin{aligned}
f=&e^{A_{34}}(a_{13}a_{23}+a_{13}a_{24}+a_{13}%
\theta_{2}+a_{14}a_{23}+a_{14}a_{24}
+a_{14}\theta_{2}+a_{23}\theta_{1}+a_{24}\theta_{1}+\theta_{1}\theta_{2}\\
&+a_{12})e^{\eta_{3}+\eta_{4}}+(a_{13}a_{23}+a_{13}\theta_{2}+a_{23}%
\theta_{1}
+\theta_{1}\theta_{2}+a_{12})e^{\eta_{3}}+(a_{14}a_{24}+a_{14}%
\theta_{2}+a_{24}\theta_{1}\\
&+\theta_{1}\theta_{2}+a_{12})e^{\eta_{4}}+\theta_{1}\theta_{2}+a_{12}\,,\\
g=&e^{A_{34}}[a_{13}a_{23}+a_{13}a_{24}+a_{13}(%
\theta_{2}+b_{2})+a_{14}a_{23}
+a_{14}a_{24}+a_{14}(\theta_{2}+b_{2})+a_{23}(\theta_{1}+b_{1})\\
&+a_{24}(\theta_{1}+b_{1})+(\theta_{1}+b_{1})(\theta_{2}+b_{2})+a_{12}]e^{%
\eta_{3}
+i\phi_{3}+\eta_{4}+i\phi_{4}}+[a_{13}a_{23}+a_{13}(\theta_{2}+b_{2})\\
&+a_{23}(\theta_{1}+b_{1})+(\theta_{1}+b_{1})(\theta_{2}+b_{2})+a_{12}]e^{%
\eta_{3}
+i\phi_{3}}+[a_{14}a_{24}+a_{14}(\theta_{2}+b_{2})+a_{24}(\theta_{1}\\
&+b_{1})+(\theta_{1}+b_{1})(\theta_{2}+b_{2})+a_{12}]e^{\eta_{4}
+i\phi_{4}}+(\theta_{1}+b_{1})(\theta_{2}+b_{2})+a_{12}, \end{aligned}
\end{equation}
Here $a_{jk}=\delta_{j}\frac{\sqrt{2}p_{j}P_{k}}{\sqrt{-P^{2}_{k}+2}+\sqrt{2}%
\delta_{k}},\delta_{j}\delta_{k}=-1, \theta_{j},\alpha_{12}$ and $%
\phi_{l},\eta_{l}\,,{A_{34}}$ are given by eqs.\eqref{N8} and \eqref{N11},
respectively. Further, we take%
\begin{equation}
\lambda_{1}=\lambda_{2}=1,P_{3}=-P_{4}=2,Q_{3}=-Q_{4}=\frac{1}{3}%
,\eta_{3}^0=\eta_{4}^0=-\frac{3\pi}{2}.
\end{equation}
thus, the corresponding semi-rational solutions $|U|$, consisting of a line rogue
wave and line breather, are obtained analytically. As can be seen in Fig.%
\ref{rb}, the first-order line rogue wave arise from the
background of first-order line breather and disappear into the same
background, and this process lasts only for a short time.
Specifically, compared to Fig.\ref{fig4} line rogue waves on the background of
the line breather do not generate high peaks(see Fig. \ref{rb}c).

\subsection{A Reduction of the reverse-space nonlocal $2$D NLS equation (%
\protect\ref{N1})}

It is universally acknowledged that $1$D nonlocal NLS equation plays an
important part in nonlinear systems. it was first introduced by Ablowitz\cite%
{Ablowitz1}, and then has been paid intensive researches on its integrable
properties and solutions\cite{Ablowitz2,llm,wxy,VS,ak,lm}. Lately, $1$D nonlocal
NLS equation was obtained in a physical application of magnetics\cite{GA}.
Subsequently, several extension visions to $1$D nonlocal NLS equation were
introduced\cite{FAS1,ab-pt}. Furthermore, $2$D NLS equation is the
high-dimensional version of $1$D NLS equation. The natural question is
whether $2$D NLS equation can reduce to $1$D NLS equation?

Setting $X=\alpha x+\beta y, T=t,$ eq. \eqref{D1} can be
reduced to the following standard $1$D reverse-space nonlocal NLS equation:
\begin{equation}  \label{41}
\begin{aligned} &iU(X,T)_{T}+\alpha \beta
U(X,T)_{XX}+\frac{\beta}{\alpha}U(X,T)U(X,T)U^{}(-X,T)=0,\\ \end{aligned}
\end{equation}
where $\alpha$ and $\beta$ are arbitrary real constants. Eq. \eqref{41} is
called the ``self-focusing" for $\alpha\beta>0$ and ``defocussing" for $%
\alpha\beta<0$, respectively. Given the solution of eq. \eqref{41}, we can
get a solution of eq. \eqref{D1}. In other words, given the solution of %
\eqref{D1}, we can get the solution of eq. \eqref{41} analytically.

\section{Reverse space-time nonlocal $2$D NLS equation \eqref{N2}}

\label{3} In this section, solutions of the reverse space-time nonlocal
eq. \eqref{N2} are considered by means of the bilinear method. Similarly, eq.%
\eqref{N2} can be rewritten as the following bilinear form:
\begin{equation}  \label{3N4}
\begin{aligned} &(D_{x}D_{y}+i D_{t})\widetilde{g} \cdot \widetilde{f} =0,\\
&(D^{2}_{x}+1)\widetilde{f} \cdot \widetilde{f} =\widetilde{g}
\widetilde{g}^{*}(-x,y,-t), \end{aligned}
\end{equation}
through the dependent variable transformation:
\begin{equation}  \label{3N5}
\begin{aligned} U=\widetilde{g}/\widetilde{f},\qquad
V=2(log\widetilde{f})_{xy}, \end{aligned}
\end{equation}
where $\widetilde{f}$ and $\widetilde{g}$ are functions with respect to
three variables $x$, $y$ and $t$, and satisfy the condition:
\begin{equation}  \label{3N6}
\begin{aligned} &V_{x}=[U(x,y,t)U^{*}(-x,y,-t)]_{y}, \quad
\widetilde{f}(x,y,t)=\widetilde{f}^{*}(-x,y,-t). \end{aligned}
\end{equation}
The operator $D$ is the Hirota's bilinear differential operator\cite{hirota}
defined by
\begin{equation}
\begin{aligned} &P(D_{x},D_{y},D_{t}, )F(x,y,t\cdot\cdot\cdot)\cdot
G(x,y,t,\cdot\cdot\cdot)\\
=&P(\partial_{x}-\partial_{x^{'}},\partial_{y}-\partial_{y^{'}},
\partial_{t}-\partial_{t^{'}},\cdot\cdot\cdot)F(x,y,t,\cdot\cdot%
\cdot)G(x^{'},y^{'},
t^{'},\cdot\cdot\cdot)|_{x^{'}=x,y^{'}=y,t^{'}=t},\nonumber \end{aligned}
\end{equation}
where $P$ is a polynomial of $D_{x}$,$D_{y}$,$D_{t},\cdot\cdot\cdot$.

The $N$-soliton solutions $U$ and $V$ of eq. \eqref{N2} can be obtained
by the bilinear transform method \cite{hirota} using \eqref{3N4} and %
\eqref{3N5}, in which $\widetilde{f}$ and $\widetilde{g}$ are written in the
following forms:
\begin{equation}  \label{3N7}
\begin{aligned}
\widetilde{f}=\sum_{\mu=0,1}\exp(\sum_{k<j}^{(N)}\mu_{k}\mu_{j}%
\widetilde{A}_{kj} +\sum_{k=1}^{N}\mu_{k}\widetilde{\eta_{k}}); \quad
\widetilde{g}=\sum_{\mu=0,1}\exp(\sum_{k<j}^{(N)}\mu_{k}\mu_{j}%
\widetilde{A_{kj}}
+\sum_{k=1}^{N}\mu_{k}(\widetilde{\eta_{k}}+i\widetilde{\Phi_{k}})),\\
\end{aligned}
\end{equation}
Here
\begin{equation}  \label{3N8}
\begin{aligned}
\widetilde{\Omega}_{j}=Q_{j}\sqrt{-P_{j}^{2}+2};\widetilde{%
\eta_{k}}=iP_{j}x+Q_{j}y +i\widetilde{\Omega}_{j}t+\eta^{0}_{j};
cos\widetilde{\Phi}_{j}=-P^{2}_{j}+1,\\
\sin\widetilde{\Phi}_{j}=\sqrt{(-P_{j}^{2}+2)}P_{j};
\exp\widetilde{(A_{jk}})=-\frac{\cos
(\widetilde{\Phi_{j}}-\widetilde{\Phi_{k}})+(P_{j}-P_{k})^2-1}{\cos(%
\widetilde{\Phi_{j}}+\widetilde{\Phi_{k}})+(P_{j}+P_{k})^2-1},\\
\end{aligned}
\end{equation}

\subsection{Periodic solutions}

\begin{figure}[!htb]
\centering
\subfigure[t=0]{\includegraphics[height=3cm,width=7cm]{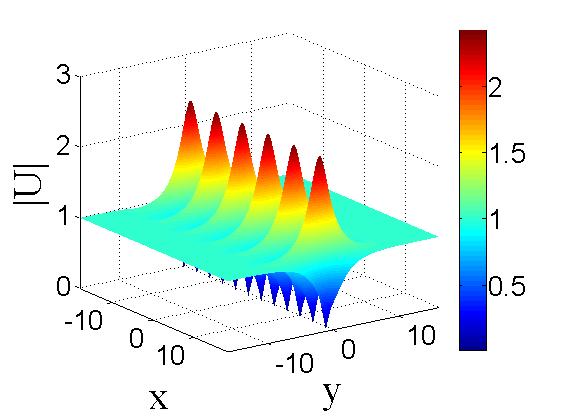}} %
\subfigure[t=0]{\includegraphics[height=3cm,width=7cm]{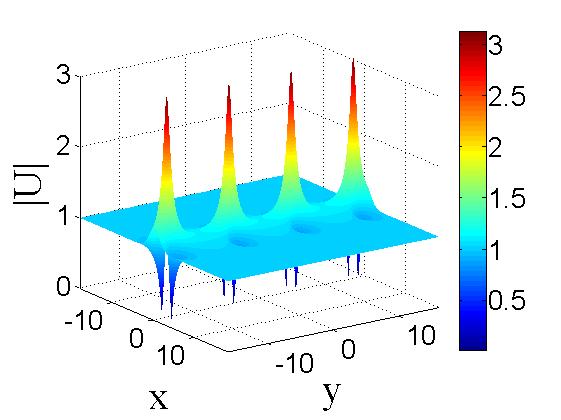}}
\caption{Two kinds of first-order breather solutions on a constant
background at time t = 0, for eq. \eqref{N2} with the parameters: (a): $%
P_{1}=\frac{1}{2},Q_{1}=1$; (b): $P_{1}=\frac{i}{2},Q_{1}=\frac{2i}{3}$.}
\label{2b}
\end{figure}

Similarly, the $nth-$order breather solutions of eq. \eqref{N2} can be
generated by taking the set of parameters in eq. \eqref{3N7}.
\begin{equation}
N=2n,P_{j}=-P_{k},Q_{j}=Q_{k},\eta _{j}^{0}=\eta _{k}^{0},  \label{3n9}
\end{equation}%
For example, we take
\begin{equation}
N=2,P_{1}=-P_{2},Q_{1}=Q_{2},\eta _{1}^{0}=\eta _{2}^{0}=0,  \label{3n91}
\end{equation}%
The first-order normal breather and Akhmediev breather solutions are derived
analytically, and its profile is shown in Fig. \ref{2b}.

\begin{figure}[!htb]
\centering
\subfigure[t=0]{\includegraphics[height=4cm,width=7cm]{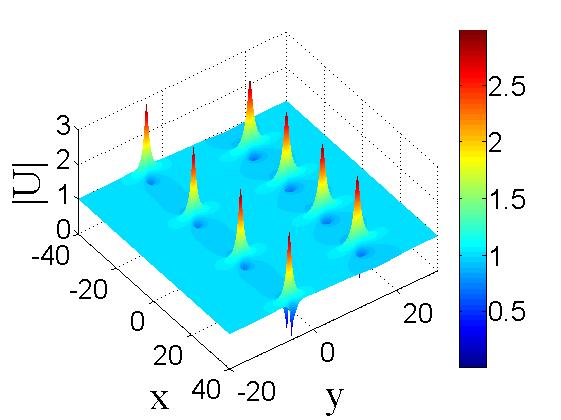}} %
\subfigure[t=0]{\includegraphics[height=4cm,width=7cm]{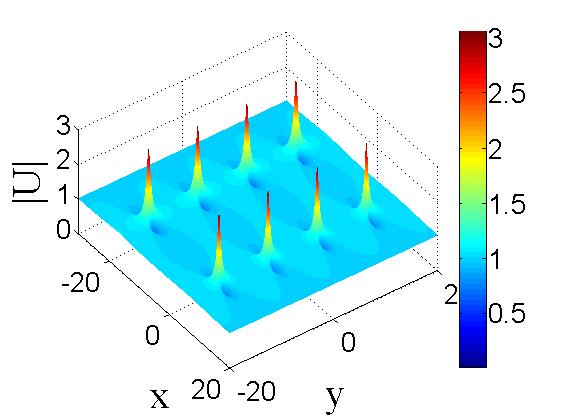}}
\caption{Two kinds of second-order breather solutions on a constant
background at time t = 0, for eq. \eqref{N2} with the parameters:(a):$%
P_{1}=\frac{1}{3},P_{3}=\frac{1}{4},Q_{1}=\frac{1}{3},Q_{3}=\frac{1}{3}$;
(b):$P_{1}=\frac{i}{3},P_{3}=\frac{i}{4},Q_{1}=\frac{2i}{3},Q_{3}=\frac{2i}{3%
}$}
\label{22b}
\end{figure}

We also give the second-order breather solutions, In this case, we set the
parameters in eq. \eqref{3N7} as follows:
\begin{equation}
N=4,P_{1}=-P_{2},P_{3}=-P_{4},Q_{1}=Q_{2},Q_{3}=Q_{4},\eta _{1}^{0}=\eta
_{2}^{0}=\eta _{3}^{0}=\eta _{4}^{0}=0,  \label{3n92}
\end{equation}
Two different forms of second-order breathers can be seen in Fig. \ref{22b}.
The higher-order breather solutions can also be generated from soliton
solutions \eqref{3N7} under parameter constraints \eqref{3n9} and They
exhibit richer dynamic behavior. We would not be here continue to
investigate this problem.

\subsection{Rational and semi-rational solutions}

The rational solutions of eq. (\ref{N2}) are generated from breathers given
by eq. (\ref{3N7}) in the long-wave limit. Taking the parameters in eq. %
\eqref{3N7} as
\begin{equation}
N=2,Q_{1}=\lambda _{1}P_{1},Q_{2}=\lambda _{2}P_{2},\lambda
_{1}=-\lambda_{2}=\lambda\neq0, \label{sN9}
\end{equation}%
and taking the limit of $P_{j}\rightarrow 0$ $(j=1,2)$, the first-order
rational solution is obtained in the following form
\begin{equation}  \label{B10}
\begin{aligned} U={\frac {2\, \left( \lambda\,y-i\sqrt {2} \right) ^{2}+2\,
\left( x- \lambda\,\sqrt {2}t \right) ^{2}+1}{2\, \left( x-\lambda\,\sqrt
{2}t \right) ^{2}+2\,{\lambda}^{2}{y}^{2}+1}} , \quad V=\,{\frac { 8 \left (
x-\lambda\,\sqrt {2}t \right) {\lambda}^{2}y}{ \left( x-\lambda\,\sqrt {2}t
\right) ^{2}+{\lambda}^{2}{y}^{2}+1/2} }. \end{aligned}
\end{equation}

The corresponding rational solutions are lumps, $(U,V)\rightarrow(1,0)$ when
$(x,y)$ goes to infinity at any given time. Lump moves along the line
direction defined by $x-\lambda\sqrt{2}t=0$, and reaches the maximum value
on this line when $y=0$. The patterns of lump solution do not change, if we
modify the time. Without loss of generality, the lump solution $U$ has the
following critical points at time $t=0$:
\begin{equation}
\Lambda_1=(0,0);\Lambda_2=(\frac{\sqrt{6}}{2},0);\Lambda_3=(-\frac{\sqrt{6}}{%
2},0);
\end{equation}

\begin{figure}[!htb]
\centering
\subfigure[$\lambda=-3$]{\includegraphics[height=3cm,width=3.7cm]{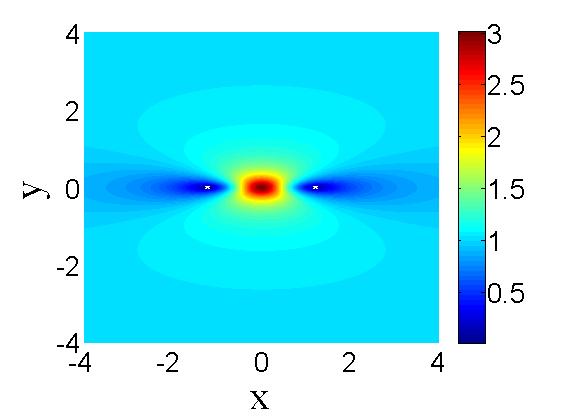}}
\subfigure[$\lambda=-1$]{\includegraphics[height=3cm,width=3.7cm]{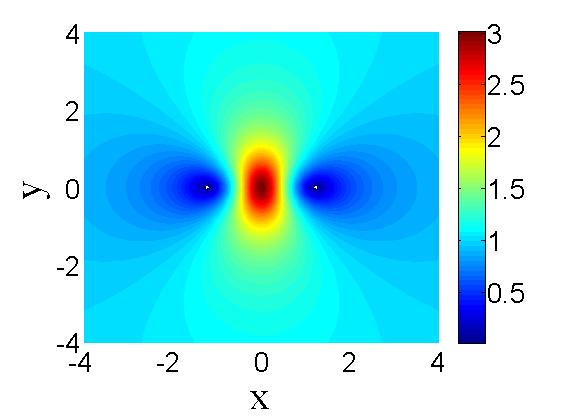}}
\subfigure[$\lambda=0.5$]{\includegraphics[height=3cm,width=3.7cm]{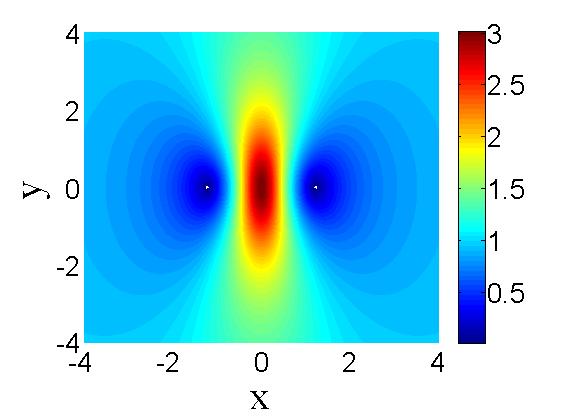}}
\subfigure[$\lambda=1.5$]{\includegraphics[height=3cm,width=3.7cm]{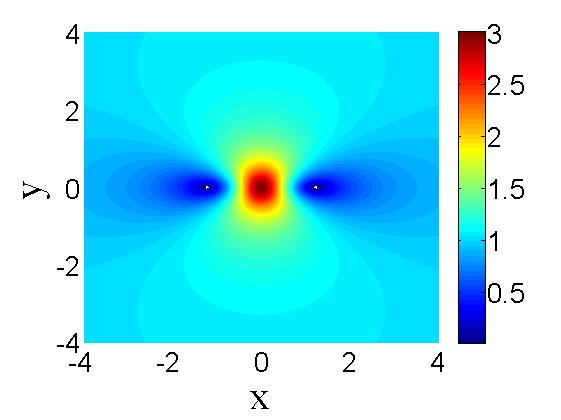}}
\caption{Lump solution $|U|$ given by Eq. \eqref{B10} in the $(x,y)-$plane at
$t=0$.}
\label{2l}
\end{figure}

Interestingly, the locality of lump is controlled by the value of $\lambda$.
As the value of $|\lambda|$ increases, the localization of the lump will be better (see Fig. %
\ref{2l}).

\begin{figure}[!htb]
\centering
\subfigure[t=0]{\includegraphics[height=4cm,width=9cm]{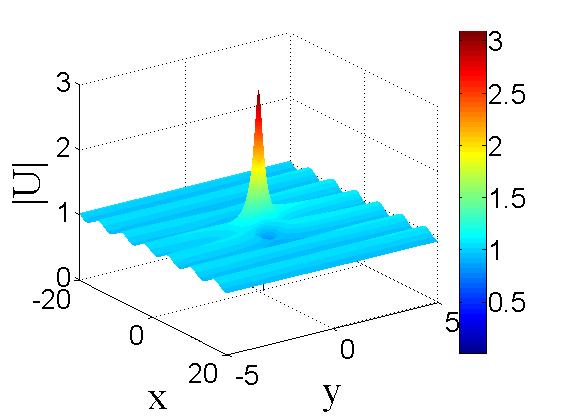}} %
\subfigure[t=0]{\includegraphics[height=4cm,width=6cm]{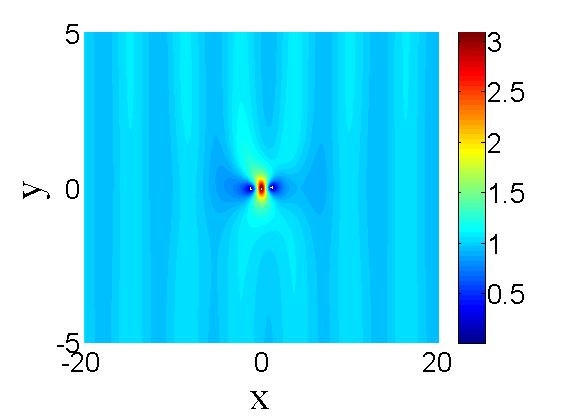}}
\caption{Semi-rational solutions $|U|$ plotted in the $(x,y)$-plane, which
consist of lump and periodic line waves with parameters given by eq.
\eqref{N17}.}
\label{lbp}
\end{figure}

For constructing semi-rational solutions of eq. \eqref{N2}, we first consider
semi-rational solutions consisting of first-order Lump and periodic line
waves obtained from the $3$-soliton solutions. For simplicity, setting
\begin{equation}  \label{sN10}
\begin{aligned}
N=3\,,Q_{1}=\lambda_{1}P_{1}\,,Q_{2}=\lambda_{2}P_{2}\,,\eta_{1}^{0}=%
\eta_{2}^{0*}=i\,\pi\,, \end{aligned}
\end{equation}
and $P_{1}\,,P_{2}\rightarrow 0$, then the functions $\widetilde{f}$ and $%
\widetilde{g}$ can be rewritten as
\begin{equation}  \label{N17}
\begin{aligned}
\widetilde{f}=&-[(x-3\sqrt{2}t+2i)^2+(3y+\sqrt{2})^2+\frac{1}{2} e^{ix-\pi}]
-(3y)^2-(x-3\sqrt{2}t)^2-\frac{1}{2},\\
\widetilde{g}=&-i[(x-3\sqrt{2}t+2i)^2+(3y+\sqrt{2}-\sqrt{2}i)^2+\frac{1}{2}
e^{ix-\pi}] -(3y-\sqrt{2}i)^2-(x-3\sqrt{2}t)^2-\frac{1}{2},\\ \end{aligned}
\end{equation}
where $\lambda_1=-\lambda_2=3,P_3=1,Q_3=0,\eta_3^0=-\pi$. This corresponding
solution $|U|$, which is plotted in Fig. \ref{lbp}, is a hybrid of a lump
and periodic line waves. The periodic line waves keep periodic in $x$
direction and localized in $y$ direction and the periodic is $2\pi$.
Compared to the height of the peak of first-order lump on a flat background
shown in Fig.\ref{2l}, an interesting phenomenon is that lump does not generate
more energy in the background of the periodic background.

\begin{figure}[!htb]
\centering
\subfigure[t=0]{\includegraphics[height=4cm,width=9cm]{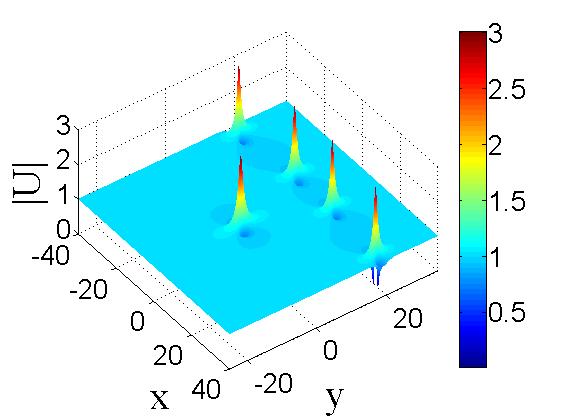}} %
\subfigure[t=0]{\includegraphics[height=4cm,width=6cm]{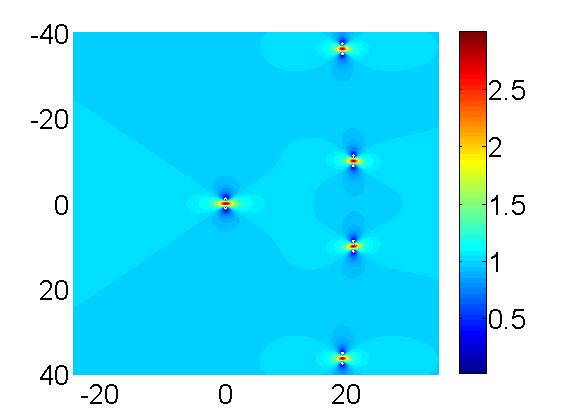}}
\caption{Semi-rational solutions $|U|$ plotted in the $(x,y)$-plane, which
consist of lump, and Akhmediev breather for eq. \eqref{N2} with parameters $%
P_3=-P_4=\frac{1}{4},Q_3=Q_4=\frac{1}{3},\protect\eta_{3}^{0}=\protect\eta%
_{4}^{0}=-2\protect\pi,\protect\eta_{5}^{0}=-2\protect\pi,\protect\lambda_1=-%
\protect\lambda_2=1$ in Eq. \eqref{N19}.}
\label{lbp1}
\end{figure}

In addition, semi-rational solutions composed of first-order lump and
Akhmediev breather can also be obtained, in a similar way, from the $4$%
-soliton solutions. We have to take the parameters
\begin{equation}  \label{N19}
\begin{aligned}
N=4\,,Q_{1}=\lambda_{1}P_{1}\,,Q_{2}=\lambda_{2}P_{2}\,,\eta_{1}^{0}=%
\eta_{2}^{0*}=i\,\pi\,,\eta_{3}^{0}=\eta_{4}^{0}, \end{aligned}
\end{equation}
and take a limit as $P_{1}\,,P_{2}\rightarrow 0$. Then the functions $%
\widetilde{f}$ and $\widetilde{g}$ can be rewritten as
\begin{equation}  \label{N20}
\begin{aligned}
\widetilde{f}&=-[A_1e^{\zeta_1}+B_1e^{\zeta_2}+C_1e^{\zeta_3}+D_1], \\
\widetilde{g}&=-[\sum_{k=1}^{3}A_{1k}e^{\zeta_1}+\sum_{k=1}^{3}C_{1k}e^{%
\zeta_3} +(A_{11}+B_{12}+A_{13})e^{\zeta_2}+D_2], \end{aligned}
\end{equation}
Here
\begin{equation}  \label{N21}
\begin{aligned}
\zeta_1&=-\frac{i}{4}x+\frac{1}{3}y-\frac{\sqrt{31}}{12}it-2\pi, \quad \zeta_2=%
\frac{i}{4}x+\frac{1}{3}y+\frac{\sqrt{31}}{12}it-2\pi, \quad
\zeta_3=\frac{2}{3}y-4\pi,\\
A_1&=[(\sqrt{2}t+x-8i)^2+(y-\sqrt{62})^2)+\frac{1}{2}],
B_1=[(\sqrt{2}t+x+8i)^2+(y-\sqrt{62})^2)+\frac{1}{2}],\\
C_1&=[\frac{32}{31}(y-2\sqrt{62})^2+\frac{32}{31}(x+\sqrt{2}t)^2+%
\frac{16}{31}], C_{12}=\frac{[(15\sqrt{31}i+97)(x+\sqrt{2}t)]^2}{124},\\
A_{11}&=\frac{[(15+\sqrt{31}i)y-16(\sqrt{62}+\sqrt{2}i)]^2}{16%
\sqrt{31}i+240}, \qquad D_2=(y+\sqrt{2}i)^2+(x-\sqrt{2}t)^2+\frac{1}{2},\\
A_{12}&=\frac{[(15\sqrt{2}+\sqrt{62}i)t+(15+\sqrt{31}i)x+8%
\sqrt{31}-120i)]^2}{16\sqrt{31}i+240},
C_{13}=\frac{1455\sqrt{31}i+1217}{1860\sqrt{31}i+12028}, \\
B_{12}&=\frac{[(15\sqrt{2}+\sqrt{62}i)t+(15+\sqrt{31}i)x-8%
\sqrt{31}+120i)]^2}{16\sqrt{31}i+240},
A_{13}=\frac{15\sqrt{31}i+9}{16\sqrt{31}i+240},\\
C_{11}&=\frac{[(15\sqrt{31}i+97)y-(883\sqrt{2}i+209\sqrt{62}i)]^2}{1860%
\sqrt{31}i+12028}, D_1=[(x+\sqrt{2}t)^2+y^2+\frac{1}{2}]. \\ \end{aligned}
\end{equation}
The corresponding semi-rational solution is illustrated in Fig.\ref{lbp1};
we see that a fundamental lump and an Akhmediev breather coexist on a
constant background, and the period of the Akhmediev breather is $8\pi$.

It is worth noting that whether the reverse space-time $2$D nonlocal NLS equation\eqref{N2} can
be reduced to the following reverse space-time $1$D nonlocal NLS equation, namely
$$iU(x,t)_{t}+AU(x,t)_{xx}+BU(x,t)U(x,t)U^{}(-x,-t)=0$$
is a meaningful issue.
We will consider it in a follow-up work.


\section{Summary}

\label{5}

In this paper, $N$-soliton solutions, periodic solutions, rational and
semi-rational solutions for the two kinds of $2$D nonlocal NLS equations
[eqs. \eqref{N1} and \eqref{N2}], which features specific $\mathcal{PT}$
symmetry with respect to different space and time, are derived by means of
the Hirota method and long-wave limit. For eq. \eqref{N1}, two types of
periodic solutions are obtained analytically. These are line breathers, and
the solutions produced by the interplay of line breathers and periodic line
waves. The first-order solutions of these two types are displayed in Figs. %
\ref{fig1} and \ref{fig2}, respectively. By taking the long-wave limit of
these periodic solutions, the W-shaped line RW (rogue-wave) and a
semi-rational solution have been constructed analytically, with examples of
them plotted in Figs. \ref{fig3} and \ref{fig4}, respectively. Moreover, the
dynamics of higher-order semi-rational solutions built of RWs and line
breathers are presented in detail (see Fig. \ref{rb}). For eq. \eqref{N2},
the obtained periodic solutions are a normal breather and Akhmediev breather
(see Figs. \ref{2b} and \ref{22b}). The nonsingular rational solution is a
lump, while the semi-rational states represent first-order lumps built on
top of the background of periodic line waves (see Figs. \ref{lbp}).
Furthermore, the corresponding semi-rational solutions consisting of the
first-order lump and Akhmediev breather are generated (see Figs. \ref{lbp1}%
). The main differences and problems for the solution of eqs.\eqref{N1} and %
\eqref{N2} can be summarized as follows:

\begin{itemize}
\item First-order breathers. Eq. \eqref{N1} produces line breathers, which
are line waves periodic in both $x$ and $y$ directions. These line breathers
disappear after a relatively short eveolution time (see Fig. \ref{fig1}). On
the other hand, eq. \eqref{N2} represents usual breathers which are periodic
in one direction and localized in the other direction, see Fig. \ref{2b}.

\item Rational solutions. The fundamental rational solution of eq. \eqref{N1}
is a line RW, whose nonvanishing amplitude persists for a very short period
of time (see Fig.\ref{fig3}). However, the nonsingular rational solution is
a lump of eq. \eqref{N2}, which is a local traveling wave keeping a constant
amplitude during its propagation, see eq.(\ref{B10}).

\item Semi-rational solutions. These solutions to eq. \eqref{N1} describe
the interaction of RWs, line breathers, and periodic line waves, see Figs. %
\ref{fig4} and \ref{rb}. The semi-rational solutions of eq. \eqref{N2} are
composed of lump, breather and periodic line waves, see Figs. \ref{lbp} and %
\ref{lbp1}.

\item The reduction problem. Eq. \eqref{N1} can be reduced to the
\textquotedblleft reverse-space" $1$D nonlocal NLS equation by a certain
transformation. However, we could not produce a $(1+1)$D reduction of the
reverse space-time $(2+1)$D nonlocal NLS equation \eqref{N2}. Thus, it is
interesting to reduce it to a reverse space-time $1$D nonlocal NLS equation.
This will be considered elsewhere.
\end{itemize}

These results essentially enrich the diversity of wave structures produced
by $2$D nonlocal NLS equations. Additionally, the distinctive structure of
these $2$D solutions is stressed by comparing them to exact solutions of the
nonlocal DS I and DS II (Davey-Stewartson) equations \cite%
{rao1,rao2,rao3,ab-pt}. These results demonstrate that the Fokas' $(2+1)$
dimensional nonlocal NLS, in the form of eqs. \eqref{N1} and \eqref{N2}, is
a valuable nonlocal extension of the NLS equation, in addition to the
previously known ones, in the form of the nonlocal DS I and DS II equations.

{Results reported in this paper provide useful models to understand
the new features of nonlinear dynamics of PT-symmetric systems \cite{KYZ1,par5},
including new types of rogue-wave solutions, which may find physical realizations
(first of all, in optics \cite{MBR1,MBR2,DR2,DR3}). Furthermore, the technique of
constructing higher-order rational and semi-rational solutions,
developed in this work, can be generalized for other nonlinear evolution equations
with $\mathcal{PT}$-symmetry.}

\section{Acknowledgments}
This work is supported by the NSF of China under Grant No. 11671219, and the
K.C. Wong Magna Fund in Ningbo University. The work of B.A.M. is partly
supported by grant No. 2015616 from the joint program in physics between the
National Science Foundation (US) and Binational Science Foundation
(US-Israel). Y. Cao and J. He thank members in their group at the Ningbo
University for useful discussions.




\section*{References}

\bibliographystyle{elsarticle-num}

\bibliography{vvv_re-modified}






\end{document}